\documentclass[twocolumn]{aastex631}

\usepackage{graphicx}
\usepackage{amsmath}
\usepackage{amssymb}
\usepackage{xspace}
\usepackage{microtype} 

\usepackage{url}
\usepackage{ragged2e}
\def\UrlBreaks{\do\/\do\-\do\.\do\=} 

\makeatletter
\def\@journalinfo{}
\long\def\frontmatter@title@above{}  
\makeatother

\shorttitle{Sub-Pixel Calibration of CMOS Sensors}
\shortauthors{Shao \& Hahn}

\begin{document}

\makeatletter
\def\footnoterule{\kern-3pt\hrule width 2in height 0.8pt\kern 8pt}
\renewcommand{\thefootnote}{}
\renewcommand{\thempfootnote}{}
\makeatother

\title{Sub-Pixel Calibration of CMOS Sensors for Precision Astrometry}

\author{M. Shao}
\author{I. Hahn}
\affiliation{Jet Propulsion Laboratory, California Institute of Technology, Pasadena, California 91109, USA}
\correspondingauthor{M. Shao}
\email{michael.shao@jpl.nasa.gov}
\footnote{\textcopyright\ 2026 California Institute of Technology. Government sponsorship acknowledged.}

\begin{abstract}
Many sources of systematic error are important to centroid the position of a star below the level of $10^{-4}$ of the FWHM of the stellar image. One of the more important sources of centroiding error is the imperfect position of the pixels in a focal plane array. In this paper, we discuss the calibration of pixel position of multiple commercial CMOS detectors. Focal plane arrays are semiconductor devices fabricated with 10's nanometer precision. However, the effective location of a pixel can be biased if there's a QE gradient across the pixel. We typically see pixel position variations of 1--3\% of a pixel in most CMOS imagers. This paper describes how we measure the position of each pixel below the $10^{-3}$ pixel level. We also describe how this information might be used in a least square fitting of an airy spot to obtain a higher accuracy position of the centroid of the PSF.
\end{abstract}


\section{Introduction}
\label{sec:intro}

When astronomers centroid star images to measure the position and motion on the sky, the normal procedure is to take the photometric value in the pixels and using an assumed point spread function (PSF) fit the photometric data to the assumed PSF, usually by allowing the total intensity and the X,Y location of the PSF to move to minimize the least square difference between the data and the model. This procedure assumes the pixels on the detector a perfectly square grid geometry. At accuracies below $\sim 1/100$ pixel the non-perfect geometry of the pixels is a dominant systematic error source (Section~\ref{sec:effect}). In this paper, we first discuss the procedure we use to measure the position of every pixel (Section~\ref{sec:pixelposition}). We discuss how the calibrated pixel position offsets can be incorporated into the least squares centroiding process (Section~\ref{sec:leastsquares}). 

This work follows our earlier work with very small format CCDs \citep{zhai2011}. Today very large format $>$ 100 Mpix exist and while astrometry with CMOS and CCDs are very similar, there subtle difference that will be discussed in some detail in Section~\ref{sec:differences}.

\section{Effect of Pixel Position on Centroid}
\label{sec:effect}

\begin{figure}[htbp]
\centering
\includegraphics[width=0.8\columnwidth]{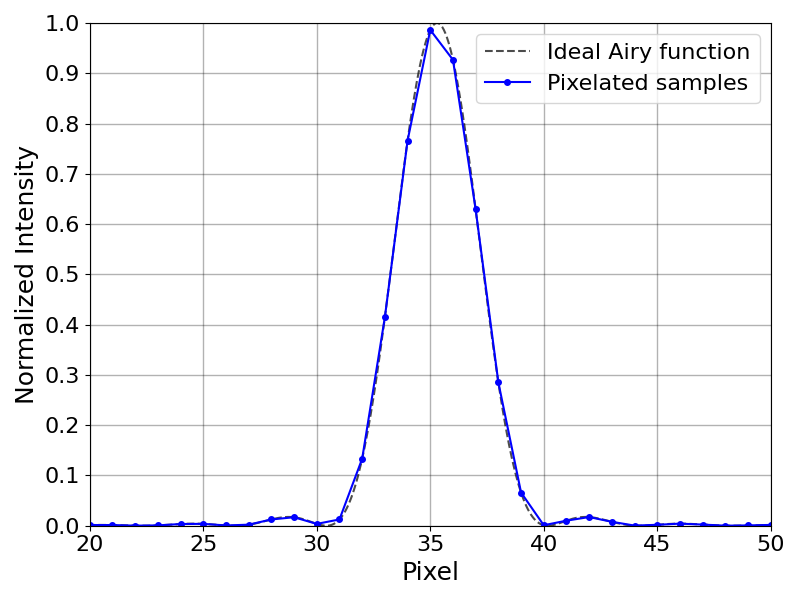}
\caption{Airy function PSF sampled at 4 pixel/$(\lambda/D)$.}
\label{fig:airy}
\end{figure}

In astronomical astrometry, one calculates the centroid of a star by performing a least-squares fit of the point spread function (PSF) to the photometric values of the pixels. For high precision astrometry it's necessary to subtract out the pixel bias as well as sky background and correct for any gain or QE difference between pixels. But usually, it's assumed that the pixels of a CCD or CMOS sensor are in a perfect square grid. But the pixel geometry is not perfect. And the position of a pixel affects the photometry mainly because the PSF has a gradient within a pixel. Figure~\ref{fig:airy} shows an airy function PSF sampled at 4 pixel/$(\lambda/D)$.

The slope at the edge of the PSF is 0.006 for a 0.01 shift in the pixel position. Or there would be a 0.6\% photometric error for a 1\% shift in the pixel position. There are several different types of pixel position errors. Many semiconductors are fabricated using a step/repeat process where there are boundary errors between steps. There can also be random errors for each pixel. In this last case we started with a perfect airy function, sampled at 4 pixels per $\lambda/D$. The analytic airy function was integrated over each pixel, using FFT convolution. To calculate the photometry of each pixel that was shifted by a random amount, we shifted the image using the FFT shift algorithm. The shift was done 4096 times for each pixel in the $64 \times 64$ pixel image. The centroid of the perturbed image was then calculated using a least square fit of an airy function. This was repeated 1000 times to calculate the rms centroiding error due to a 1\% random pixel position offset for every pixel. Because the centroid is calculated using many pixels, one would expect the rms error to be smaller than 0.01 pixel. The result of the simulation was that a detector with random pixel position error of 0.01 pixel the centroid of a PSF sampled with 4pix/$\lambda/D$ is 0.00158 pixels.
\section{Measuring Pixel Position Offsets}
\label{sec:pixelposition}

Attempts to measure imperfections in the geometry of CCD devices goes several decades \citep{buffington1990}. The technique we used to measure the position of every pixel follows our earlier work \citep{zhai2011}. We collaborated with F. Malbet at U. Grenoble whose team subsequently continued this research \citep{crouzier2016,lizzana2024,rousset2026}. The basic idea is to illuminate the detector with the interference pattern produced by two coherent single-mode optical fibers. Since each fiber emits a nearly ideal spherical wave, their superposition forms a sinusoidal fringe pattern with well-defined spatial frequency and phase. If one of the optical paths is phase-modulated---for example with a piezoelectric fiber stretcher or an electro-optic phase modulator---the fringes sweep smoothly across the detector by a known fraction of a fringe period.

Each pixel records a sinusoidal intensity variation as the fringes move. The phase of this modulation depends on the exact position of the pixel center relative to the fringe pattern, while the modulation amplitude provides information about the pixel response. By fitting the temporal signal from every pixel, one measures the phase at each pixel with extremely high precision.

To recover the full two-dimensional detector geometry, the measurement is repeated with fringes at two or more different orientations. A horizontal fringe pattern measures the pixel coordinates perpendicular to the fringes (e.g., the x-coordinate), while a vertical or oblique pattern measures the orthogonal direction. Combining these measurements yields a dense map of the actual pixel-center positions relative to an ideal rectangular grid. The resulting distortion map captures lithographic placement errors, local stretching or compression of the silicon lattice, stitching errors between reticle fields, and other manufacturing imperfections that cannot be determined from flat fields or stellar observations alone. During subsequent astrometric reductions, stellar centroid measurements are corrected using this calibrated pixel geometry, removing a systematic error source that otherwise becomes dominant once photon noise is reduced below roughly $10^{-2}$ pixel.

\begin{figure}[htbp]
\centering
\includegraphics[width=\columnwidth]{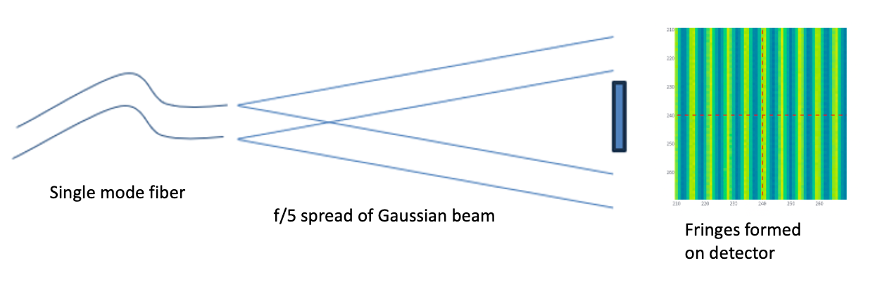}
\caption{A basic principle of the pixel position calibration using single mode fibers.}
\label{fig:principle}
\end{figure}

One of the strengths of the interferometric approach is that it measures geometry independently of the point spread function (PSF). Traditional centroiding with stars or pinhole masks is always convolved with optical aberrations and PSF modeling errors, whereas the fringe phase is referenced to the optical wavelength and depends almost entirely on the physical location of the detector sampling points. Because billions of photons can be accumulated over long integrations, the phase of the fringe can be determined to a tiny fraction of a cycle. For visible wavelengths, and a fringe spacing of $\sim$6.3 pixels per fringe cycle, a phase precision of $10^{-4}$ radians correspond to $\sim$350 picometers on a detector with 3.5 $\mu$m pixels, roughly $10^{-4}$ pixel in effective position.

\subsection{Description of the setup and the data taking procedure}

\begin{figure}[htbp]
\centering
\includegraphics[width=\columnwidth]{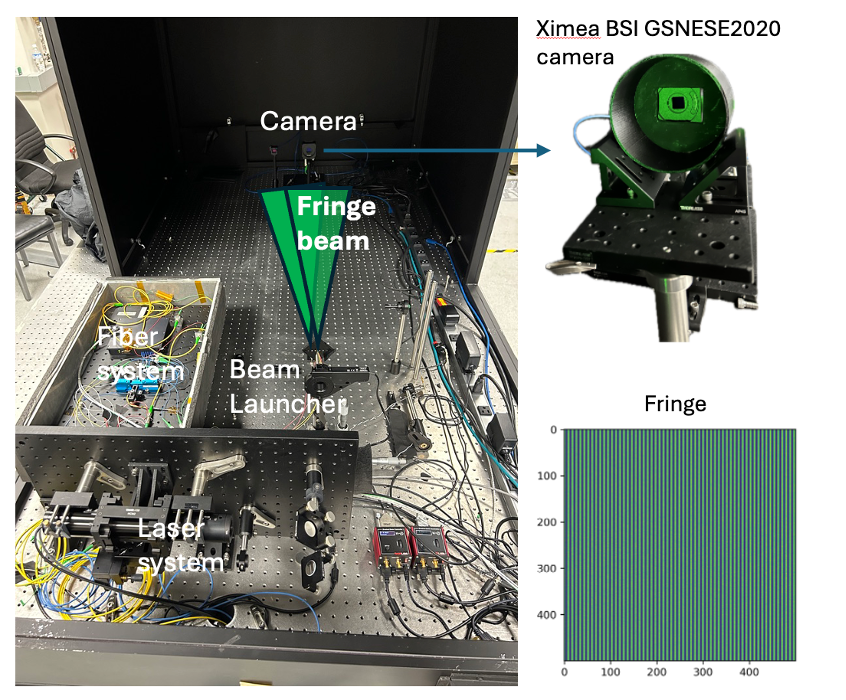}
\caption{A photo of a detector calibration hardware setup. A XIMEA sCMOS sensor is used for the test. The beam launcher is rotatable to form an interference fringe pattern at any angle with respect to the detector coordinate. A typical vertical direction fringe pattern image (500$\times$500 ROI) is shown above.}
\label{fig:setup}
\end{figure}

The detector calibration setup is housed inside a light-tight enclosure on an optical table. A photo of the setup is shown in Figure~\ref{fig:setup}. The fiber system sub-bench is inside a thermally controlled box to minimize the temperature-dependent changes in the optical path difference (OPD). Two different laser systems (533nm and 633nm) have been used for the calibration, although the photo shows only the He-Ne (633nm) laser system. The beam launcher is mounted on a rotation stage to form vertical and horizontal fringes with respect to the detector's coordinate system. A Python script was developed to control the experiments and acquire the calibration frames from the camera. The calibration frames involve dark frames, two flat-field frames (flat1, flat2), and fringe frames. A pixel-to-pixel Quantum Efficiency (QE) variation map is independently measured using a non-coherent LED light source with an integrating sphere. This resulting QE map is used to correct the individual pixel's photon responsivity to a 0.1\% level. The experiment script first measures the darks and then takes flat1 and flat2 measurements (1500 frames each), followed by the fringe measurement (typically 1500 frames). The calibration run is repeated to reduce photon noise and to check long-term stability of the measurements. The Region of Interest (ROI) is typically set to a 400$\times$400 pixel area near the center of the detector. The limited ROI, rather than the full frame, was chosen to speed up data processing time. Camera electronics typically outputs a combined output from two ADCs (low and high gain). To simplify the analysis of the camera readout circuits nonlinearity, only low gain ADC chain is used to acquire the image. To obtain the pixel-error map, the fringe data must be acquired at different optical phases. In other words, every individual pixel must see photon fluxes covering a full amplitude of an ideal sinusoidal optical wave to calculate pixel-size variations. The phase of the optical sine wave (i.e. the optical fringe pattern) is modulated by a fiber stretcher using a function generator.

\begin{figure}[htbp]
\centering
\includegraphics[width=0.7\columnwidth]{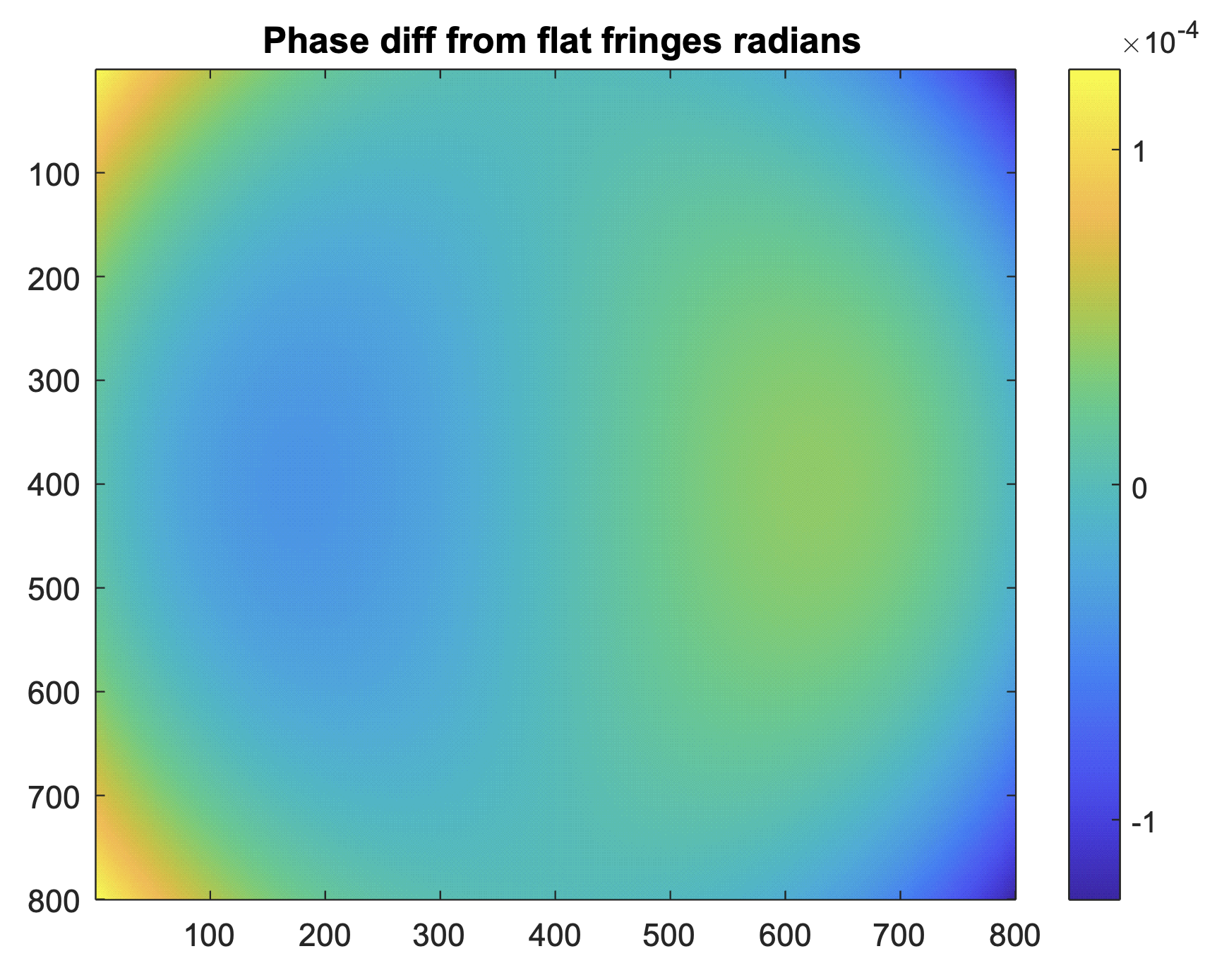}
\caption{Phase bias estimation from the hyperbolic fringe effect in our setup.}
\label{fig:hyperbolic}
\end{figure}

\subsection{Description of analysis, phase of a pixel relative to other pixels}

The metrology processing algorithm works as follows: First, the dark level of each pixel is computed from darks run (no external light). Next, spatial fringe fitting (using a two-dimensional cosine function) is performed on every frame. The darks-subtracted fringe frame is corrected by subtracting (flat1+ flat2) and normalizing it by the QE variation pixel map. In this step, pixel positions are assumed to be perfectly regular, represented by $(i,j)$. An initial guess of the spatial frequency parameter is obtained by performing a 2D FFT on the corrected fringe frame. The pixel-size variation determination assumes that the measurement fringe beam is optically perfect, meaning a single cosine wave can be parametrized by five parameters (offset, amplitude, wavenumbers $(k_x,k_y)$, and phase). The hyperbolicity of the fringe, which results from combining non-planar, spherical wavefronts from the fiber tips, is estimated to be negligible. Once the spatial fringe fit of all frames is complete, pixel location deviations (dx and dy) from the ideal, integer mesh pattern $(i,j)$ are computed to determine 2D pixel location variation map. A mathematical representation of the variation map calculation is shown in Equation~\ref{fig:math}.

\begin{figure*}[htbp]
\centering
\small

\begin{equation}
\begin{aligned}
&\text{1) Spatial fringe fit} \\
&\quad I_n^{i,j} = I_n^0 + A_n^0 \cos\left(\phi_n^0 + \vec{\kappa}_n^0 \cdot \vec{r}_{i,j}\right) \quad \text{where} \quad \vec{\kappa}_n^0 \cdot \vec{r}_{i,j} = \kappa_{n,x}^0 \cdot j + \kappa_{n,y}^0 \cdot i \\
&\quad \xrightarrow[\text{fit all pixels for each frame}]{\text{non-linear least squares}} \{I_n^0, A_n^0, K_{n,x}^0, K_{n,y}^0, \phi_n^0\} \\
&\text{2) Pixel locations calculation} \\
&\quad I_n^{i,j} = I_n^0 \iota_{i,j} + A_n^0 \alpha_{i,j} \cos\left(\phi_n^0 + \vec{\kappa} \cdot \vec{r'}_{i,j}\right) \\
&\quad = I_n^0 \iota_{i,j} + A_n^0 \cos\phi_n^0 \cdot \underbrace{\alpha_{i,j} \cos\left(\vec{\kappa} \cdot \vec{r'}_{i,j}\right)}_{C_{i,j}} - A_n^0 \sin\phi_n^0 \cdot \underbrace{\alpha_{i,j} \sin\left(\vec{\kappa} \cdot \vec{r'}_{i,j}\right)}_{S_{i,j}} \\
&\quad \text{where} \quad \vec{\kappa} \equiv \langle \vec{\kappa}_n^0 \rangle \quad \vec{r'}_{i,j} = (i,j) + \hat{\kappa}\delta r_{i,j} \quad \text{and} \quad \hat{\kappa} \equiv \vec{\kappa}/|\vec{\kappa}| \\
&\quad \begin{bmatrix} \boldsymbol{\iota}_{i,j} \\ \mathbf{C}_{i,j} \\ \mathbf{S}_{i,j} \end{bmatrix} = \begin{bmatrix}
\mathbf{I}_1^0 & \mathbf{A}_1^0 \cos\phi_1^0 & -\mathbf{A}_1^0 \sin\phi_1^0 \\
\mathbf{I}_2^0 & \mathbf{A}_2^0 \cos\phi_2^0 & -\mathbf{A}_2^0 \sin\phi_2^0 \\
\vdots & \vdots & \vdots \\
\mathbf{I}_N^0 & \mathbf{A}_n^0 \cos\phi_N^0 & -\mathbf{A}_N^0 \sin\phi_N^0
\end{bmatrix}^\dagger \begin{bmatrix} \mathbf{I}_1^{i,j} \\ \mathbf{I}_2^{i,j} \\ \vdots \\ \mathbf{I}_N^{i,j} \end{bmatrix} \\
&\quad \alpha_{i,j} = \sqrt{C_{i,j}^2 + S_{i,j}^2} \\
&\quad \vec{\kappa} \cdot \vec{r'}_{i,j} = \tan^{-1}\frac{S_{i,j}}{C_{i,j}} + m2\pi \rightarrow \delta r_{i,j} = \frac{1}{|\vec{\kappa}|}\left(\tan^{-1}\frac{S_{i,j}}{C_{i,j}} + m2\pi - \vec{\kappa} \cdot (i,j)\right) \text{ along } \hat{\kappa}
\end{aligned}
\label{fig:math}
\end{equation}

\end{figure*}

\vspace{1em}

\subsection{2$^{\text{nd}}$ level effects: hyperbolic fringes and orthogonality of axes}

The fringes from two optical fibers to the eye look like perfectly straight fringes. Perfectly straight fringes are the result of two plane wavefronts interfering. But the wavefronts from the two fibers are spherical and strictly speaking the fringes are hyperbolas. For our experimental setup the curvature of the fringes is very small. To quantify the effect of hyperbolic fringes we simulated a setup like our current lab setup.

The two fibers were separated by $\sim$2.5cm and the distance to the detector was $\sim$1.5m. The laser wavelength was 532nm and the pixel size 3.45$\mu$m. The fringe period is approximately 10 pixels. When measuring the position of the pixels, there will be a systematic offset because the fringes are hyperbolic rather than straight lines. The bias varies across the detector as shown in Figure~\ref{fig:hyperbolic}. The color scale is in radians of phase, $10^{-4}$ radians are $1.7 \times 10^{-4}$ pixels or $\sim$0.59 nm on the x,y position of the pixel detector. Our goal for this phase of the work is $10^{-3}$ pixels.

For both dX and dY measurements of each pixel position, we need to put both horizontal and vertical fringes on the detector. The fringes don't have to be exactly parallel to either the rows or columns of pixels but ideally the two sets of fringes need to be perpendicular to each other. If they are not perpendicular, there will be what we call a parallelogram error. For $10^{-3}$ pixel accuracy over a 400$\times$400 pixel field of view the non-orthogonal error should be $< 2.4$ $\mu$rad.

\subsection{Results from Gpixel and HWK sensors}

\subsubsection{Gpixel sensor calibration result}

\begin{figure*}[htbp]
\centering
\includegraphics[width=0.9\textwidth]{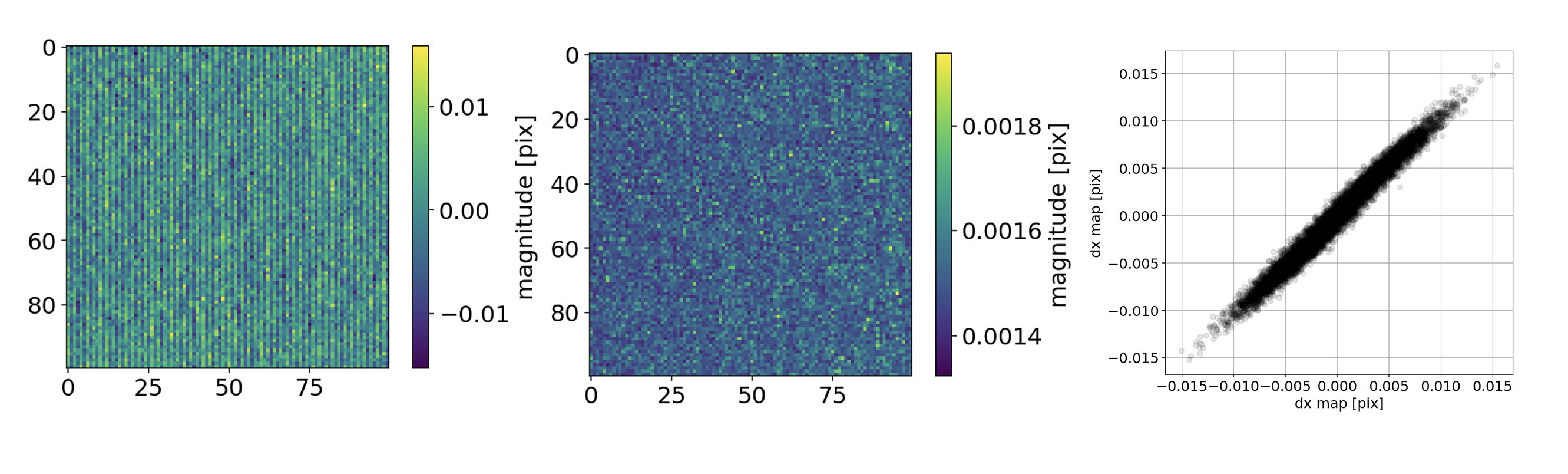}
\caption{Pixel position variation map of a Gpixel sensor in x-direction (left), and its standard deviation map (middle). The position variation map consists of a mean of 2000 measurements. The standard deviation map shows that the precision of measurement is currently limited by the photon noise ($\sim$0.15\%). The long-term stability is verified by comparing with measurements taken a month later. The pixel-to-pixel correlation plot of two dx-maps measured in different time are shown (right).}
\label{fig:gpixel_dx}
\end{figure*}

Figure~\ref{fig:gpixel_dx} shows the pixel location variation map in x-axis. Only dX-map is shown for clarity. The map revealed a non-random alternating pattern in row direction which is a typical design feature of 2-layer stacked CMOS sensor architecture where pixels are structurally mirrored or grouped to share peripheral readout circuitry. The vertical deep contacts that line the top photodiode layer to the bottom transistor follow an alternating arrangement across the pixel grid. The dY-map shows no alternating pattern as expected.

\begin{figure*}[htbp]
\centering
\includegraphics[width=0.9\textwidth]{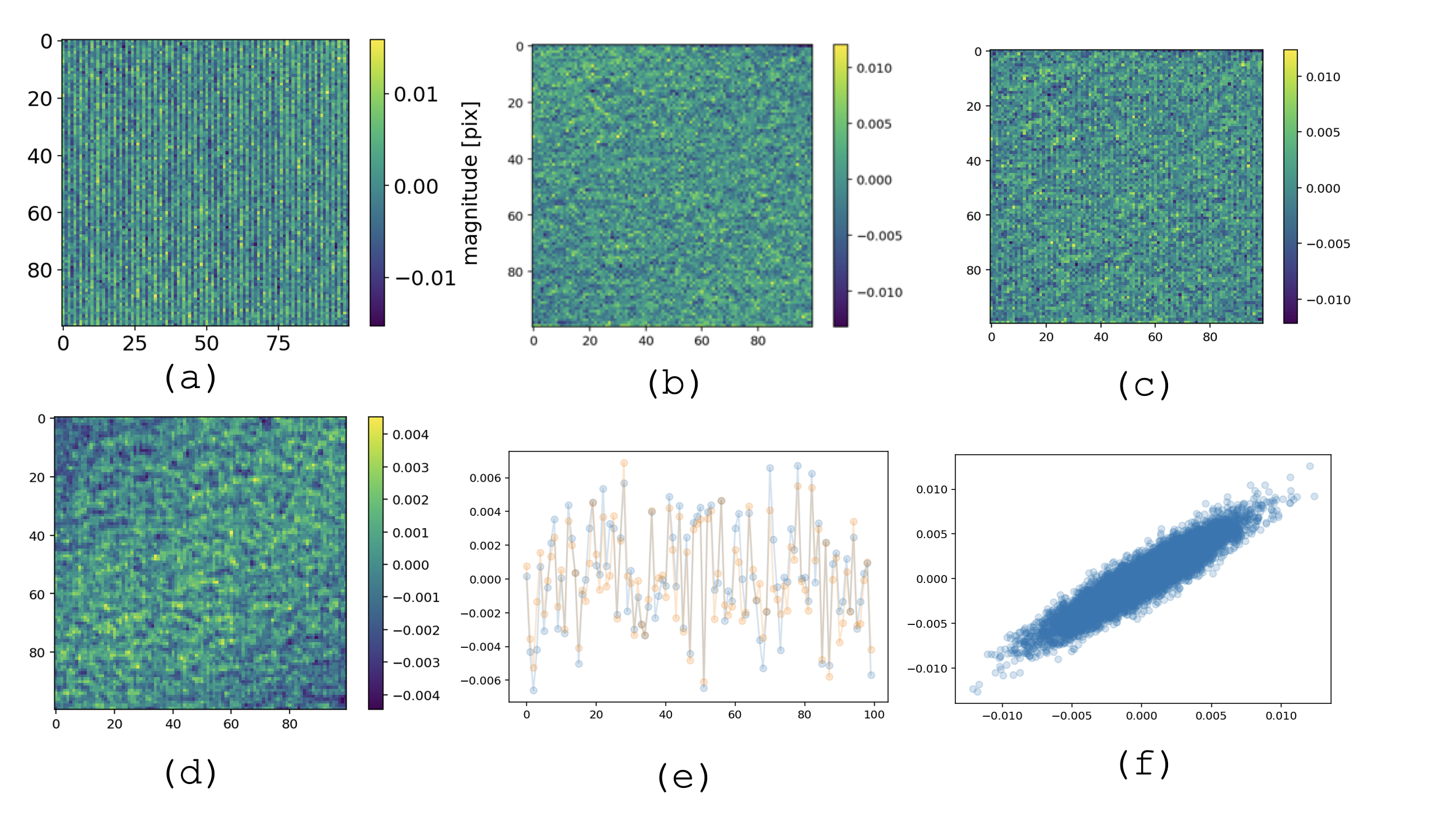}
\caption{Accuracy verification of dX (a) and dY (b) maps using a diagonal dXY map (c). (d) is the difference map calculated using the geometric relationship: diff=$dx \times \cos(45^\circ)+dy \times \cos(45^\circ)-dXY$. (e) it shows a good correlation between the calculated dXY (blue) and the measured dXY(orange) along the same row. (f) correlation between the measured and the calculated dXY maps using all pixels. The measurements is currently limited by photon noise at 0.15\% rms level.}
\label{fig:gpixel_verification}
\end{figure*}

The repeatability of the calibration map is check by comparing the same dX-map taken a month later. The correlation between the maps is only limited by the photon noise. We also verified the accuracy of the dX and dY maps using its geometrical relationship with a variation map measured along a diagonal direction of the chip. Figure~\ref{fig:gpixel_verification} shows the dX, dY and dXY (diagonal) direction pixel variation maps and their correlation.

\subsubsection{HWK sensor calibration result}

\begin{figure*}[htbp]
\centering
\includegraphics[width=0.7\textwidth]{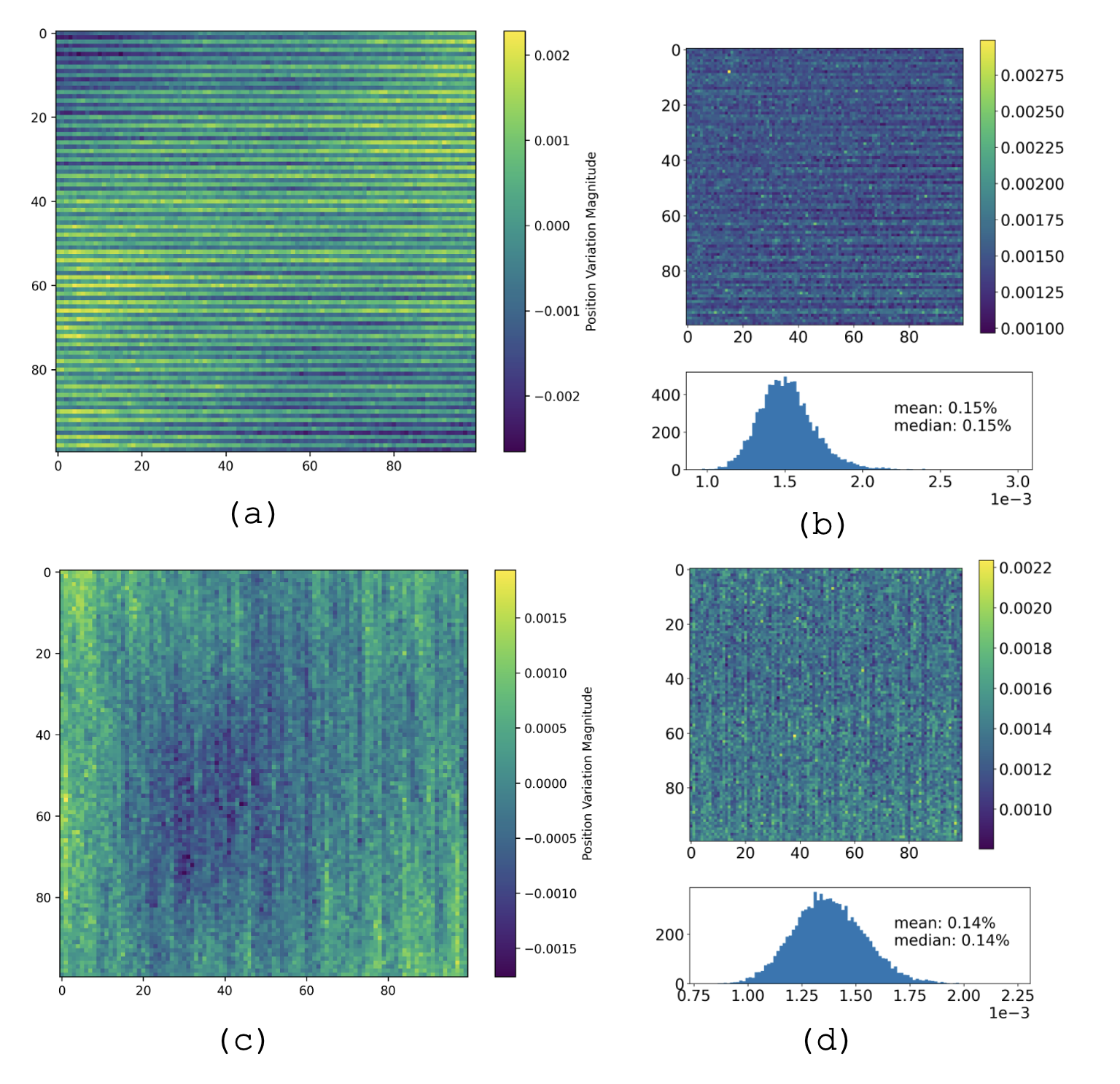}
\caption{Position variation maps of a HWK CMOS sensor: (a) x-direction map and (b) the standard deviation map. (c) y-direction map and (d) the standard deviation map.}
\label{fig:hwk}
\end{figure*}

Figure~\ref{fig:hwk} shows the pixel location variation map maps and the precision of the measurement. The x-direction map shows the 2-layered stacked CMOS architecture of modern backside illuminated sensor architecture.

\begin{figure*}[htbp]
\centering
\includegraphics[width=0.75\textwidth]{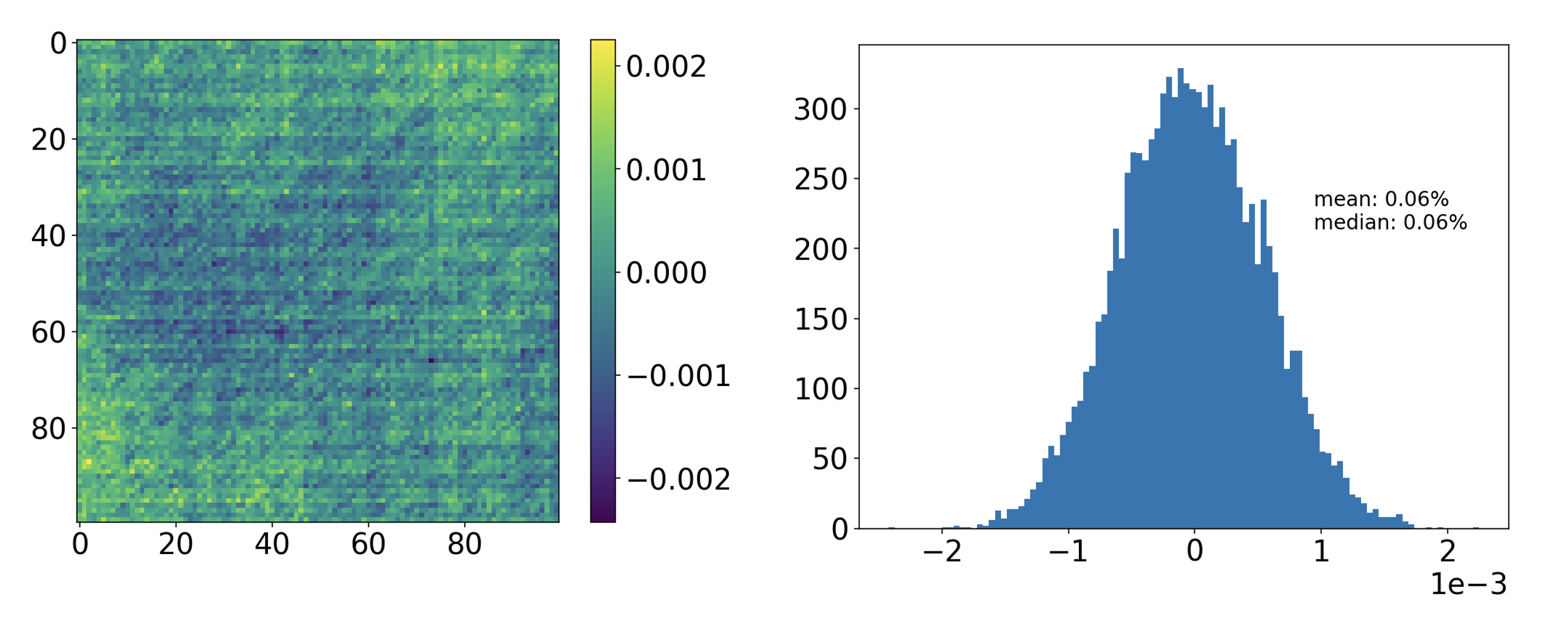}
\caption{Repeatability and accuracy of the maps: (left) difference between the measured 45 degree-direction map (dXY) and calculated from dX and dY maps using simple geometrical relationship. The map measurements were taken in different time separated by a few days. (right) the histogram of the difference map shows a good repeatability and the limitation of the current calibration of the detector, $\sim$0.06\% pixel.}
\label{fig:hwk_accuracy}
\end{figure*}

Figure~\ref{fig:hwk_accuracy} represents the accuracy of the map. To check the accuracy, we measured a variation in 45-degree direction (dXY map) by forming a diagonal fringe on the detector. This measured map is then compared with a calculated map using dX and dY maps using the assumed squareness of the detector over the region of interest. Measurements to generate each map were performed in different days. The difference between the calculated dXY map and the measured one represents the accuracy of dX and dY maps.

\section{Differences CCD vs CMOS}
\label{sec:differences}

Several types of CMOS focal planes cannot be calibrated. In this section we discuss which types can't and why.

\subsection{Microlenslet array}

Most CMOS sensors have microlenslet arrays. For front sided CMOS focal planes, the photosensitive area of a pixel is significantly smaller than the pixel spacing. This leads to very low quantum efficiency. Microlenslet arrays are put on top of the silicon to increase effective collecting area of every pixel.

\begin{figure}[htbp]
\centering
\includegraphics[width=0.6\columnwidth]{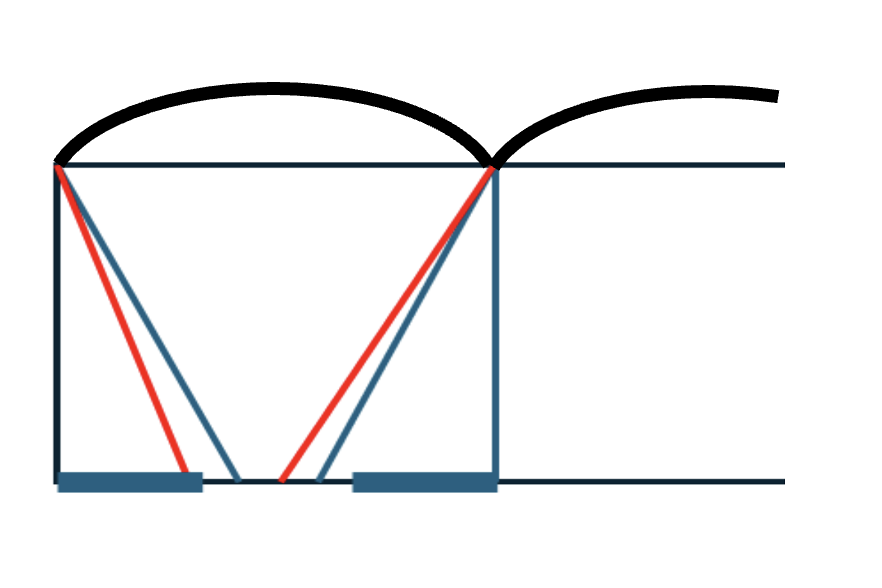}
\caption{An illustration of complexity of calibrating detectors with microlenslet arrays. The microlenslet are simple lenses and the focus may not be perfect. The photometric center of the pixel changes depending on the angle of incidence of the light.}
\label{fig:microlens}
\end{figure}

The lenslet problem for the sub-pixel detector position calibration is schematically illustrated in Figure~\ref{fig:microlens}. Light that hits the detector on axis, all of the light is intercepted by the phot-responsive part of the pixel. But light that enters off axis light from one side of the pixel hits a non-photoresponsive part of the detector. As a result, the photocenter of the pixel changes depending on the angle of incidence of the light. Instead of measuring the X, Y photocenter of each pixel, we must measure X,Y as a function of $\theta$-$\phi$ the angle the light is hitting the detector. For practical purposes we call this type of sensor uncalibratable.

One might think that lenslet are not needed in a backside CMOS focal plane array. But in practice a great many backside CMOS sensors use lenslet arrays. This includes several CMOS sensors by Sony, which are otherwise excellent devices.

\subsection{Rolling vs Global Shutter}

When taking images of a star field, the telescope pointing will not be perfect, especially at the $10^{-3}$ to $10^{-4}$ pixel level. The motion of the telescope will cause the image to look like the convolution of the motion and the diffraction limited image. If there is a bright star in the FOV, the motion within one exposure can be measured and this modified PSF can be used to analyze the centroid. But this is true only if the camera has a global shutter. With a rolling shutter the exposure time of each row of pixels is shifted in time relative to the previous row. If the telescope moved in a non-linear way the PSF at the top of the image can be slightly different than at the bottom of the image. The same effect can produce errors when we're projecting laser fringes onto the detector. In the case of pixel position calibration, we take many images to average down this source of error.

\subsection{Other differences}

The list at this point is a list of effects that don't introduce significant errors in calibrating the XY geometry of a CMOS detector, as long as the analysis of the data takes them into account. CMOS detectors have a separate read amp for each pixel and a separate A/D converter for each column of pixels. The digital output of the sensor will have a bias (the digital value in the absence of light), gain, QE that is different for each pixel. Many CMOS detectors have very low read noise and moderately large full well.  Often there will be two A/D converters a low gain A/D that saturates at 50,000e and a high gain A/D that has $<$ 1.5e read noise for low level signals. Often the camera has an FPGA that blends two 11 bit or 12 bit A/Ds into a single 16 bit output. Almost all the time the ``blending'' will introduce systematic errors make pixel position calibration below $10^{-3}$ pixels impossible. Many CMOS detectors give the user the option to use just one of the A/D outputs.  Even then we have observed significant non-linearity in sensor of $\sim$1.5\%. When we analyze the fringe pattern in the data, we do a FFT to get the spatial frequency $(k_x, k_y)$ of the fringe. Nonlinearity shows up as a 2nd harmonic in the FFT spectrum.

\subsection{Etaloning effect}

In coherent laser-based pixel metrology using interferometry, etaloning can produce interference patterns when coherent light undergoes multiple reflections between parallel or near-parallel surfaces, such as the layers within an image sensor or its cover glass. This effect introduces measurement artifacts and reduces accuracy. For the accuracy required at the $10^{-3}$ pixel level, we found that the cover glass must be removed. We also found that thin-layered detectors are difficult to calibrate due to etaloning. The penetration depth in silicon (a typical material used in BSI CMOS sensors) depends strongly on the laser wavelength. By selecting an appropriate laser wavelength, the sensor's etaloning effect can be minimized.

\section{Least Squares Centroid Using Pixel Position Offsets}
\label{sec:leastsquares}

There are several different ways the known imperfect pixel geometry of the sensor can be incorporated into the least-square fitting process. First, we review traditional approach where the pixel geometry is assumed perfect, then extend the centroid analysis procedure to include pixel position offsets. There are many ways to incorporate pixel position offsets what's illustrated here is not the only one.

In high accuracy astrometry one starts with a model of the PSF of the telescope. This might be an analytical model such as an airy function with the central obscuration and spider in the model. If the PSF is Nyquist sampled, we can use a numerical model. The model will have 3 adjustable parameters, x, y, intensity. These three parameters are adjusted until the rms difference between the model and the data is minimized. In differential astrometry with real telescopes, the PSF will vary across the FOV. But in differential astrometry where one might be measuring the parallax of a star or the wobble of a star due to an orbiting planet, we're watching the same part of the sky over time. The motions are usually quite small. One approach might be to cross correlate a star's position at one epoch with its position at a different epoch. The exact pixel location would change between epochs simply because the telescope pointing might be a few arcsec or 10's of arcsec shifted between epochs. The image at one epoch can be shifted a fraction of a pixel using the Fourier shift theorem until the rms difference is minimized.

When the pixel geometry is imperfect, we have two problems. If we look at the image at epoch 1, the photometric value at each pixel is slightly wrong. Because of this error, even with Nyquist sampling, we don't have the true PSF. Similarly with the image of the same star at a different epoch but on a different set of pixels. If we use FFT to shift one image, we will have an error, that'll be large compared to $10^{-3}$ pixels.

One solution is making use of the fact that the pixel position error is usually small a few \% of the width of the pixel. With the imperfect image, we can FFT shift the image at each pixel according to the pixel position map measured interferometrically. Alternatively, we use the FFT to measure the gradient of the photometry when the pixel is shifted, for example 1\% in X and 1\% in Y. We use this gradient to correct for the photometric error. After correcting for this position error, we now have the ``true'' PSF, still Nyquist sampled. Similarly, we calculate the ``true'' PSF for the same star at a different epoch landing on a different set of pixels. We can then measure the motion of the star between epoch with a least-square fit by FFT shifting one image relative to the other.

\section{Conclusion and Discussion}
\label{sec:conclusion}

One effect that deserves some discussion is stray laser light hitting the detector. When two beams of laser light interfere, the fringe visibility is $2\sqrt{A_1 \cdot A_2}/(A_1+A_2)$, where $A_\#$ is the intensity of the light. Another way to put this is, that the amplitude of the E fields add/subtract, not their intensity. As a result, scattered laser light with an intensity of $10^{-6}$ can affect the phase at $10^{-3}$ radians which is near the $10^{-3}$ pixel accuracy we're aiming for in this stage of the work. In particular, light in the cladding in the single mode fiber needs to be eliminated. Another consideration is that this experiment was conducted in air. There was an enclosure to prevent either warm or cold air from the building's HVAC from directly blowing into the optical path. And we took a lot of data which averaged out much of the turbulence. But beyond $10^{-3}$ pixels the measurements are much easier in a vacuum chamber. We have shown repeated measurements of the pixel positions of these CMOS devices over a month are repeatable below $5 \times 10^{-4}$ pixels rms. And a check on the accuracy by the measurements by comparing dX and dY measurements with a measurement with the fringes at 45deg are consistent at the $\sim 10^{-3}$ pixel level need for the future SHERA mission.

\begin{acknowledgments}
This research was carried out at Jet Propulsion Laboratory, California Institute of Technology, under a contract with the National Aeronautics and Space Adiministration (80NM0018D0004). We would like to thank Nathan Bush for setting up the HWK camera set up, and Gautam Vashisht for discussions regarding the SHERA mission concept and its requirements on the detector calibration.
\end{acknowledgments}

\begingroup
\setlength{\bibindent}{0pt}
\footnotesize
\small
\begingroup
\setlength{\bibindent}{0pt}

\makeatletter
\g@addto@macro{\UrlBreaks}{\UrlOrds}
\makeatother
\setlength{\Urlmuskip}{0mu plus 1mu} 

\let\oldsetup\thebibliography
\renewcommand{\thebibliography}[1]{%
  \oldsetup{#1}%
  \setlength{\itemsep}{0pt plus 0pt minus 0pt}  
  \setlength{\parsep}{0pt plus 0pt minus 0pt} 
  \setlength{\topsep}{0pt plus 0pt minus 0pt}   
  \justifying                                   
  \sloppy                                       
}

\vfill
\endgroup

\end{document}